\documentclass[onecolumn]{elsart3p}

\usepackage{graphicx}
\usepackage{dcolumn}
\usepackage{bm}
\usepackage{amssymb}

\begin{document}

\begin{frontmatter}

\title{Stability of an $[N/2]$-dimensional invariant torus
in the Kuramoto model at small coupling}

\author{Hayato Chiba}
\address{Department of Applied Mathematics and Physics,
Kyoto University, Kyoto, 606-8501, Japan}

\author{Diego Paz\'o}
\address{Instituto de F\'{\i}sica de Cantabria,
IFCA (CSIC-UC), E-39005 Santander, Spain}

\date{\today}

\begin{abstract}
When the natural frequencies are allocated symmetrically 
in the Kuramoto model there exists an invariant
torus of dimension $[N/2]+1$ ($N$ is the population size).
A global phase shift invariance allows to reduce the model to 
$N-1$ dimensions using the phase differences,
and doing so the invariant torus becomes $[N/2]$-dimensional. 
By means of perturbative calculations based on the renormalization group technique, we show that
this torus is asymptotically stable at small coupling if $N$ is odd. If $N$ is even the torus can be stable
or unstable depending on the natural frequencies, and both possibilities
persist in the small coupling limit.
\end{abstract}


\end{frontmatter}

\section{Introduction}
The Kuramoto model \cite{Kuramoto,strogatz00,rmp_kuramoto} has become the basic
framework for the description of macroscopic synchronization; 
a phenomenon observed in a variety of natural and artificial
systems~\cite{PRK,MMZ04}.
Kuramoto \cite{Kuramoto} considered a population of all-to-all weakly coupled
oscillators
such that their interaction could be reduced to their phases:
\begin{equation}
\dot \theta_j = \omega_j + \frac{\varepsilon}{N} 
\sum_{l=1}^N f( \theta_l- \theta_j),\quad  j=1,2, \cdots ,N,
\label{kuramoto_model}
\end{equation}
where $\theta_j$ and $\omega_j$ are, respectively, the phase and the natural
frequency
of the $j$-th oscillator, and $\varepsilon$ is the coupling strength.
Kuramoto adopted a sinusoidal coupling function $f(\cdot)=\sin(\cdot)$ together
with
a symmetric frequency distribution of the natural frequencies,
what resulted very useful for the theoretical analysis of the model.

Originally, it was useful and instructive to consider the thermodynamic limit of
(\ref{kuramoto_model}),
$N\to\infty$. Finite-size effects have remained unsolved for a long time and
only recently significative
advances have been achieved
\cite{daido90,balmforth2000,hong07,hildebrand07,buice07}.
Also some attention has been recently devoted to the small-$N$ behavior of the
Kuramoto model
by Maistrenko and coworkers from the point of view of dynamical systems
theory \cite{maistrenko04,maistrenko05,maistrenko05_ijbc}.

In this paper we study the Kuramoto model with a finite population,
and with the natural frequencies allocated symmetrically 
around the mean frequency. 
One of the reasons that motivates this problem is the fact that most works on
the Kuramoto model
have assumed that the natural frequencies are distributed according
to a symmetric probability density,
and as a consequence it is usual that numerical simulations are carried out 
selecting frequencies not at random, but reflecting the inherent symmetry of the
frequency distribution.
In particular, several
works~\cite{maistrenko04,maistrenko05,maistrenko05_ijbc,popovych05}
have recently investigated phase diagrams of the Kuramoto model with a finite
population $N$
under the assumption that the natural frequencies are allocated symmetrically.
It has been shown that under these assumptions, finite $N$ and symmetry of the
natural frequencies,
model (\ref{kuramoto_model}) exhibits a peculiar type of chaos dubbed `phase
chaos'.

Of more importance for this work is the finding in \cite{popovych05} that 
the phase space contains an $[N/2]$-dimensional invariant torus.
This torus has been thought to be unstable ({\em i.e.}~repelling) 
when $\varepsilon\to0$ \cite{maistrenko05,maistrenko05_ijbc,popovych05}.
This belief is probably motivated by the difficulty of
investigating numerically the phase diagram for small $\varepsilon$
due to the extremely weak (in)stability of invariant sets in that limit.
One of the purposes of this paper is to reveal the phase diagrams of the
Kuramoto model
at small coupling and different values of $N$. Our analytical results are
obtained
by using the renormalization group (RG) method,
which is one of the singular perturbation methods.
Our results firmly establish the stability of 
the mentioned invariant torus on rigorous mathematical grounds.
In particular, we give general results for any finite $N$, and some small
populations ($N=3, ..,7$)
are investigated in more detail.

\section{Basic definitions}

Due to the mean-field character of the Kuramoto model we may arbitrarily label
the natural frequencies from the smaller to the larger: $\omega_1 \leq
\omega_{2} \leq \cdots \leq \omega_N$.
Moreover, by going into a suitable rotating framework we set the mean frequency
equal to zero
without loss of generality. For the discussion to follow it is worth to note
that
if there are not coincident natural frequencies ({\em i.e.} $\omega_1<\omega_2<
\cdots < \omega_N$),
some degree of synchronization, $\left<\dot\theta_i\right> =
\left<\dot\theta_j\right>$ for some (or all) $i\ne j$,
is only achieved for a coupling strength 
larger than some positive constant $\varepsilon_c$.

Next, we rewrite the Kuramoto model for the convenience of our analysis.
Due to the invariance of the global quantity $\Theta=\sum_{l=1}^N \theta_l$,
the Kuramoto model can be reduced in one dimension by changing 
to a new set of coordinates:
$\varphi_j=\theta_{j+1}-\theta_{j}, \,j=1,\cdots ,N-1$. 
It is also useful to define ``frequency gaps" $\Delta_j=\omega_{j+1}-\omega_{j},
\,j=1,\cdots ,N-1$. 
In the new variables the Kuramoto model has this structure
\begin{equation}
\dot{\varphi }_j = \Delta_j + \varepsilon \, \Xi_j(\varphi_1,\varphi_2,
\ldots, \varphi_{N-1}) , \quad j= 1,\cdots ,N-1. 
\label{kuramoto_model2}
\end{equation}

Maistrenko and coworkers~\cite{maistrenko05,maistrenko05_ijbc,popovych05} found
that if the natural frequencies are symmetrically selected
($\omega_i=-\omega_{N-i+1}$ $\Rightarrow$ $\Delta_j=\Delta_{N-j}$) the Kuramoto
model has  an invariant manifold $\mathcal{M}$, namely a torus of dimension
$[N/2]$, which is given by $\mathcal{M}=\{\varphi_i=\varphi_{N-i}\}$.
(In the original coordinates, $\mathcal{M}$ is any of the tori
that ---parameterized by  the invariant $\Theta$--- foliate
an $([N/2]+1)$-dimensional torus; thus if one selects $\Theta =0$,
$\mathcal{M}=\{\theta_i= - \theta _{N-i+1}\}$.)

\section{Main results}

In Refs.~\cite{maistrenko05,maistrenko05_ijbc,popovych05} 
it was shown that for some values of the parameters $\omega _i$ and $\varepsilon $,
the dynamics approaches the invariant torus $\mathcal{M}$ or an attractor $\mathcal{A} \subset \mathcal{M}$.
They reported that the regions of parameters with stable $\mathcal{M}$ or $\mathcal{A}$
are close to, or inside, synchronization regions 
(what implies that $\varepsilon$ is larger than some
positive number if $\omega_i\ne\omega_{j\ne i}$).
Thus, there is a common belief that far from synchronization,
as the coupling strength goes to zero ($\varepsilon \to 0$),
the dynamics fills the whole phase space, the $N-1$ dimensional torus $\{\varphi_j\}$.
So far, many numerical results confirmed this expectation
\cite{maistrenko05,maistrenko05_ijbc,popovych05}. However, we show in this paper 
that this is not true.
We have found that in the limit $\varepsilon\to 0$:

\begin{enumerate}
\item[(i)] If $N$ is odd the $(N-1)/2$-dimensional invariant torus $\mathcal{M}$
is, unless there are special resonant conditions among the natural frequencies, asymptotically stable.
\item[(ii)] If $N$ is even the $N/2$-dimensional invariant torus $\mathcal{M}$
can be stable or unstable depending on the particular disposition of the natural frequencies.
\end{enumerate}

Statements (i) and (ii) are consequence of three Theorems
to be proved by using the RG method, which is a powerful singular
perturbation method for differential equations proposed in \cite{CGO1,CGO2}.
Recently, mathematical foundation of the RG method was given in \cite{Chiba1,Chiba2}
showing that the RG method is useful as well to investigate existence and stability
of invariant manifolds. We present a brief review of the 
RG method in Section 7.1, and the proofs of the main theorems in this paper
are shown in Section 7.2 and 7.3.

\section{Odd $N$}

In this section, we investigate the stability of the invariant torus $\mathcal{M}$ for odd $N$.
For two particular cases, $N=3$ and $5$, the phase diagrams are completely uncovered.

One of the main theorems in this paper is as follows:\\

\textbf{Theorem 1.} \,\,\, Suppose that $N = 2M-1$ is an odd number.
If the natural frequencies satisfy the following nonresonance condition:
\begin{eqnarray}
\left\{ \begin{array}{l}
\omega _i = \omega _j \,\,\, \mathrm{if \, and \, only \, if}\,\,\, i=j, \\
\omega _k + \omega _j = 2\omega _i \,\,\, \mathrm{if \, and \, only \, if}\,\,\, i=k=j \,\,\mathrm{or}\,\,  j=2M-k, i=M, \\
\omega _i + \omega _j = \omega _k + \omega _l \,\,\, \mathrm{if \, and \, only \, if}\,\,\, 
i=j=k=l \,\, \mathrm{or} \,\, j=2M-i, l=2M-k, \\
3\omega _i = \omega _j + \omega _k + \omega _l \,\,\, \mathrm{if \, and \, only \, if}\,\,\, 
i=j=k=l, \\
\omega _i + 2\omega _k = \omega _j + 2\omega _l \,\,\, \mathrm{if \, and \, only \, if}\,\,\, 
i=j, k=l \,\, \mathrm{or} \,\,j=2M-i, k=M, l=i,
\end{array} \right.
\label{NRcondition}
\end{eqnarray}
then there exists a positive constant $\varepsilon _0$, which depends on the natural frequencies, 
such that if $0<\varepsilon <\varepsilon _0$, the invariant torus $\mathcal{M}$ is asymptotically stable
and the transverse Lyapunov exponents of $\mathcal{M}$ are of $O(\varepsilon ^3)$. \\

Equation~(\ref{NRcondition}) can be rewritten as a condition for $\Delta _j$'s by using the relation
$\omega _j = -\sum^{M-1}_{k=j}\Delta_k$.
The proof of this theorem is given in Sec.~7.2.

As $\varepsilon\to 0$, $[N/2]$-frequency quasiperiodic dynamics on $\mathcal{M}$ is stable for almost all $\{ \Delta _i\}$.
Parameter regions on which $\varphi _i$'s are (partially) phase-locked on 
$\mathcal{M}$ are very narrow.
On such regions, there exist a $k$-dimensional stable torus on $\mathcal{M}$ 
filled by $k$-frequency quasiperiodic orbits ($k < [N/2]$). 
Further, we can prove that regions with phase-locking are
narrower when the nonresonance condition is fulfilled 
than when it is not.

Below we test the validity of Theorem 1 for $N=3$, 5, and 7.
Focusing on particular cases will allow us to understand better how Theorem 1 
applies in practical terms. For instance, 
for $N=5$ we make a complete analysis of the stability of $\mathcal{M}$,
showing what happens when the natural frequencies do not satisfy the nonresonance condition of Theorem 1.

\subsection{$N=3$}

For $N=3$ it is particularly simple to prove the stability of $\mathcal{M}$ using basic theory of dynamical systems.

The ODEs ruling the dynamics in $\{\varphi_i\}$ coordinates [Eq.~(\ref{kuramoto_model2})] are:
\begin{equation}
\left\{ \begin{array}{l}
\displaystyle \dot{\varphi }_1 = \Delta + \frac{\varepsilon }{N} 
  \left[ \sin \varphi _2 - 2\sin \varphi _1 - \sin (\varphi _1 + \varphi _2) \right],   \\
\displaystyle \dot{\varphi }_2 = \Delta + \frac{\varepsilon }{N} 
  \left[ \sin \varphi _1 - 2\sin \varphi _2 - \sin (\varphi _1 + \varphi _2) \right],  \\
\end{array} \right.
\label{N3}
\end{equation}
where we are already assuming the symmetry $\Delta_1=\Delta_2\equiv\Delta$.
The dynamics inside the invariant 1-torus $\mathcal{M}$ ({\em i.e.}~a circle defined by $\varphi_1=\varphi_2\equiv\varphi$) obeys
\begin{equation}
\dot\varphi = \Delta - \frac{\varepsilon}{3}[\sin \varphi + \sin(2\varphi)] 
\label{N3-2}
\end{equation}
and a transverse perturbation $\delta\varphi \equiv \varphi_2 - \varphi_1$ is governed by 
\begin{equation}
\dot{\delta\varphi} = - \varepsilon \cos \varphi  \, \delta\varphi + O(\delta\varphi^2) .
\end{equation}
The transverse Lyapunov exponent (TLE) is hence:
\begin{equation}
\lambda_\perp = - \varepsilon \int_0^{2\pi} P(\varphi) \cos \varphi  \, d\varphi ,
\end{equation}
with $P(\varphi)= C  / \dot\varphi$ for $\varepsilon$ smaller than the synchronization threshold $\varepsilon_c\approx 1.704 \, \Delta$. $C$ is a normalization constant such that $\int_0^{2\pi} P(\varphi) \, d\varphi=1$.
Making an expansion of $P(\varphi)$ in terms of the small
quantity $\varepsilon/\Delta$, it turns out that $\lambda_\perp$ becomes negative with a
cubic dependence on $\varepsilon$:
\begin{equation}
\lambda_\perp=-\frac{1}{18}\frac{\varepsilon^3}{\Delta^2}\left[1+\frac{1}{6}\left(\frac{\varepsilon}{\Delta}\right)^2 +
 O\left(\left(\frac{\varepsilon}{\Delta}\right)^4\right) \right] .
\label{n3}
\end{equation}

This result agrees\footnote{The nonresonance condition is not fulfilled if and only if
$\Delta=0$, and in that case the TLE
has a linear dependence on $\varepsilon$: $\lambda_\perp=-\varepsilon$.
Indeed, it is easy to see that the fixed point $\varphi =0$ of Eq.~(\ref{N3-2}) is stable for any $\varepsilon >0$.} 
with Theorem 1.
Finally, it must be noted that the invariant torus exists for any odd interacting function $f(\varphi)=-f(-\varphi)$ 
and not only for the particular choice $f(\varphi)=\sin(\varphi)$.
The transverse Lyapunov exponent is then:
\begin{equation}
\lambda_\perp=-\frac{1}{18\pi}\frac{\varepsilon^3}{\Delta^2} \int_0^{2\pi} f'(\varphi) 
\left[ 2f(\varphi)f(2\varphi)+ f^2(2\varphi)\right] \, d\varphi +O(\varepsilon^5) ,
\label{f}
\end{equation}
with $f'(\varphi)\equiv df(\varphi)/d\varphi$.

\subsection{$N=5$}

If $N=5$, Eq.~(\ref{kuramoto_model2}) reads
\begin{equation}
\left\{ \begin{array}{l}
\displaystyle \dot{\varphi }_1 = \Delta_1 + \frac{\varepsilon }{N} 
     \Bigl( -2\sin \varphi _1  + \sin \varphi _2 - \sin (\varphi _1 + \varphi _2) + \sin (\varphi _2 + \varphi _3) \\
     \quad \quad - \sin (\varphi _1 + \varphi _2 + \varphi _3) + \sin (\varphi _2 + \varphi _3 + \varphi _4)
                 - \sin (\varphi _1 + \varphi _2 + \varphi _3 + \varphi _4)  \Bigr),   \\
\displaystyle \dot{\varphi }_2 = \Delta_2 + \frac{\varepsilon }{N} 
     \Bigl( -2 \sin \varphi _2 + \sin \varphi _1 + \sin \varphi _3  \\
     \quad \quad - \sin (\varphi _1 + \varphi _2) - \sin (\varphi _2 + \varphi _3) + \sin (\varphi _3 + \varphi _4)
                 - \sin (\varphi _2 + \varphi _3 + \varphi _4)  \Bigr),   \\
\displaystyle \dot{\varphi }_3 = \Delta_2 + \frac{\varepsilon }{N} 
     \Bigl( -2 \sin \varphi _3 + \sin \varphi _4 + \sin \varphi _2  \\
     \quad \quad - \sin (\varphi _3 + \varphi _4) - \sin (\varphi _2 + \varphi _3) + \sin (\varphi _1 + \varphi _2)
                 - \sin (\varphi _1 + \varphi _2 + \varphi _3)  \Bigr),   \\
\displaystyle \dot{\varphi }_4 = \Delta_1 + \frac{\varepsilon }{N} 
     \Bigl( -2\sin \varphi _4  + \sin \varphi _3 - \sin (\varphi _3 + \varphi _4) + \sin (\varphi _2 + \varphi _3) \\
     \quad \quad - \sin (\varphi _2 + \varphi _3 + \varphi _4) + \sin (\varphi _1 + \varphi _2 + \varphi _3)
                 - \sin (\varphi _1 + \varphi _2 + \varphi _3 + \varphi _4)  \Bigr).
\end{array} \right.
\label{N5}
\end{equation}
Again, symmetry is assumed ($\Delta_1=\Delta_4$, $\Delta_2=\Delta_3$), and hence
there is an invariant 2-torus 
$\mathcal{M} = \{ \varphi _1 = \varphi _4 ,\, \varphi _2 = \varphi _3\}$.
Since $\mathcal{M}$ is $2$-dimensional, dynamics on $\mathcal{M}$ is nontrivial.
Depending on values of $\Delta_1, \Delta_2$ and $\varepsilon$, the asymptotic dynamics on $\mathcal{M}$
can be quasiperiodic or periodic (fixed points only exist 
above a finite $\varepsilon$ value unless $\Delta_1=\Delta_2=0$).
Quasiperiodic motion is generic when $\varepsilon \to 0$.
Periodic motion exist inside open sets in the phase diagram (so-called Arnold tongues) whose widths 
shrink to zero as $\varepsilon \to 0$.
Inside an Arnold tongue there is (at least) one pair of stable-unstable (along the torus surface) periodic orbits whose
average frequencies are related by a rational rotation number:
$\left< \dot \varphi_1 \right> : \left< \dot \varphi_2 \right> = n:m$.
We call a periodic orbit of that type an $n:m$ {\em locking solution}.
Arnold tongues touch the axis $\varepsilon=0$ at the points where the ratio $\Delta _1 : \Delta _2$ 
is rational. Major Arnold tongues are born at (rational) $\Delta _1 : \Delta _2$ ratios 
corresponding to frequencies that do not fulfill the nonresonance condition of Theorem 1.

In what follows, we assume\footnote{In the degenerate case $\Delta _2 = 0$
the nonresonance condition of Th.~1
is violated if and only if $\omega _1 = \omega _2 = 0$
($\Delta _1 = \Delta _2 = 0$).
In this case, Eq.~(\ref{N5}) has a stable fixed point $\varphi _i=0,\, i= 1,\ldots,4$
with TLEs $\lambda_\perp^{(1)}=\lambda_\perp^{(2)}=-\varepsilon$.}
$\Delta _2 \neq 0$. By rescaling time and $\varepsilon$, 
we are allowed to divide Eq.~(\ref{N5}) by $\Delta _2$ to assume that $\Delta _2 = 1$
without loss of generality.
Then, the nonresonance condition for $N=5$ gives $\Delta _1 \neq 0, 1/2, 1,2,3,4$.
We can prove the next theorem.\\

\textbf{Theorem 2.} There exists a non-negative number $\varepsilon _0 = \varepsilon _0(\Delta_1)$ 
such that the invariant torus $\mathcal{M}$ is asymptotically stable if $0 < \varepsilon < \varepsilon _0$.
$\varepsilon _0(\Delta_1)$ tends to zero as $\Delta_1 \to 1/2, 1,2, 3$. \\

\begin{figure}
\begin{center}
\includegraphics[]{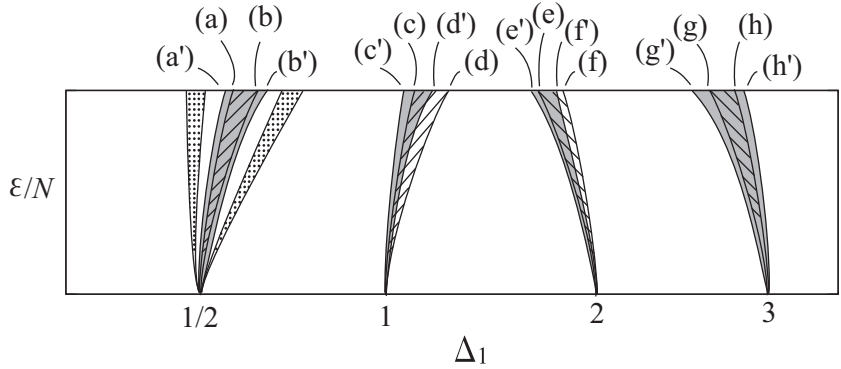}
\caption[]{A schematic view of the phase diagram for $N=5$ and small $\varepsilon $. 
Inside the hatched regions $\mathcal{M}$ is unstable; and
the dotted regions intend to represent finitely many 
disjoint hatched regions.
Shaded regions are Arnold tongues with rotation numbers 
$n:m = 1:2,\, 1:1,\, 2:1$ and $3:1$ from left to right.}
\label{tongue_n5}
\end{center}
\end{figure}

A sketch of the strategy for proving this theorem is given in Sec.~7.3. And in Sec.~7.4 we show
explicitly how the proof yields the phase diagram near $\Delta_1=1/2$
(the most intricate case).
We note that Th.~2 asserts that 
the cases $\Delta_1 = 0$ and $4$ do not yield transversal instability of $\mathcal{M}$ despite
of violating the nonresonance condition of Th.~1.
A schematic view of the phase diagram of Eq.~(\ref{N5}) for small $\varepsilon $ is represented in Fig.~\ref{tongue_n5},
in which the invariant torus $\mathcal{M}$ is unstable in the tongue-shaped hatched regions.
The $n:m$ locking solutions for  $n:m = 1:2,\, 1:1,\, 2:1$ and $3:1$ exist in the gray regions.
In particular, in the gray regions that are not hatched there are transversally stable $n:m$ locking solutions.
Asymptotic expansions with respect to $\varepsilon /N$ of boundaries (a) to (h) and (a') to (h') in Fig.~\ref{tongue_n5}
are shown in the Appendix (the expansion for each line is done up to an order
that completely unfolds the phase diagram).
In two dotted regions emerging from $\Delta_1= 1/2$, there are many disjoint unstable tongue-shaped regions,
while exactly one unstable region emerges from each of the other resonances:
$\Delta_1 = 1,2,3$.
The existence of such many unstable regions emerging from $\Delta_1=1/2$ 
is shown in Section 7.3 resorting to the RG method and with aid of numerical simulations.

\begin{figure}
\begin{center}
\includegraphics[width=5.0in]{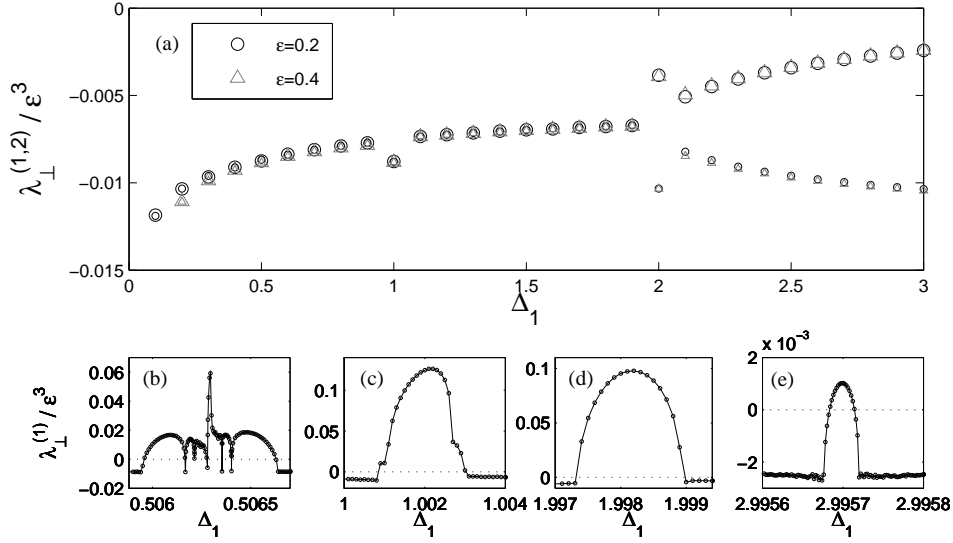}
\caption[]{(a) Transverse Lyapunov exponents scaled by $\varepsilon^3$ 
as a function of $\Delta_1$ for $\varepsilon=0.2,0.4$ ($N=5$, $\Delta_2=1$).
(b-e) Regions where the largest TLE becomes positive. (It is instructive 
to compare the panel (b) with the result in Fig.~\ref{app:numerics}(a)
obtained by the RG method.)}
\label{num_n5}
\end{center}
\end{figure}

\begin{figure}
\begin{center}
\includegraphics[]{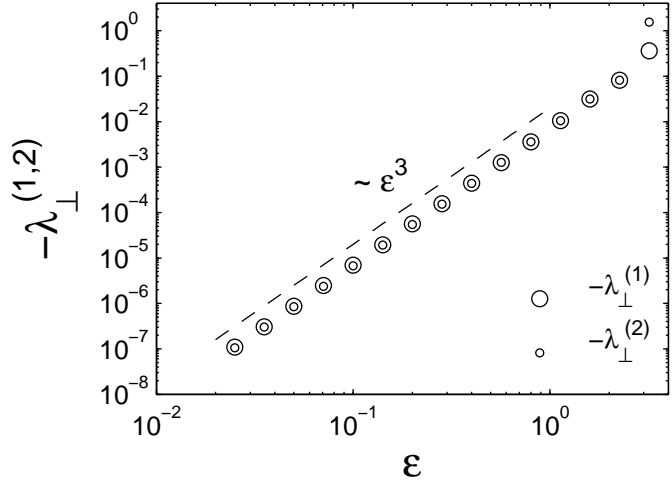}
\caption[]{TLEs as a function of $\varepsilon $ for $N=5, \Delta _2 = 1$ and $\Delta _1 = (1 + \sqrt{5})/2$ (the
golden mean).}
\label{n5_e3}
\end{center}
\end{figure}

Next, we present numerical results corroborating our theoretical results for $N=5$.
The dynamics of infinitesimal perturbations transversal to the torus $(\delta\varphi_1\equiv\varphi_1-\varphi_4, \delta\varphi_2\equiv\varphi_2-\varphi_3)$ is governed by two linear equations:
\begin{equation}
\left\{ \begin{array}{l}
\displaystyle \dot{\delta\varphi_1}=-\frac{\varepsilon}{N}\cos(\varphi_1+\varphi_2)(1+4\cos\varphi_2)\delta\varphi_1 + \frac{\varepsilon}{N}\left[\cos\varphi_2 -\cos(\varphi_1+\varphi_2)\right]\delta\varphi_2 ,\\
\displaystyle \dot{\delta\varphi_2}=-\frac{\varepsilon}{N}\cos(\varphi_1+\varphi_2)\sin^2(\varphi_2/2) \delta\varphi_1 
-\frac{\varepsilon}{N}\left[3\cos\varphi_2 +2\cos(\varphi_1+\varphi_2))\right]\delta\varphi_2 .
\end{array} \right.
\label{dn5}
\end{equation}
From these equations we calculate the
TLEs ($\lambda_\perp^{(1)} \ge \lambda_\perp^{(2)}$)
using the popular method by Benettin {\em et al.}~\cite{benettin80}.
In Fig.~\ref{num_n5}(a) we plot the TLEs 
for two values of $\varepsilon$ observing that: (i) TLEs are almost everywhere negative 
as expected from Theorem 1,
(ii) TLEs for $\varepsilon = 0.2$ and $0.4$ collapse when divided by $\varepsilon^3$,
except if one enters in a locking region nor fulfilling the nonresonance condition in Th.~1
($0:1$ locking in the leftmost part of the panel).
Figure \ref{n5_e3} shows a log-log plot of the TLEs as a function of $\varepsilon$
for a specific value of $\Delta_1$ (arbitrarily chosen 
with the constraint that the nonresonance condition is satisfied).
We find a nice power law $\lambda_\perp^{(1,2)}=-|\lambda_\perp^{(1,2)}|\propto \varepsilon^3$,
as expected from Theorem 1.
In Figs.~\ref{num_n5}(b-e) we depict the largest TLE in four regions about
lockings $1:2$, $1:1$, $2:1$, and $3:1$, finding that it becomes positive in short intervals
as advanced in Theorem 2.
These intervals match with the analytical expressions in Appendix.

\subsection{$N= 7$}
\label{sec_n7}

\begin{figure}
\begin{center}
\includegraphics[width=5.0in]{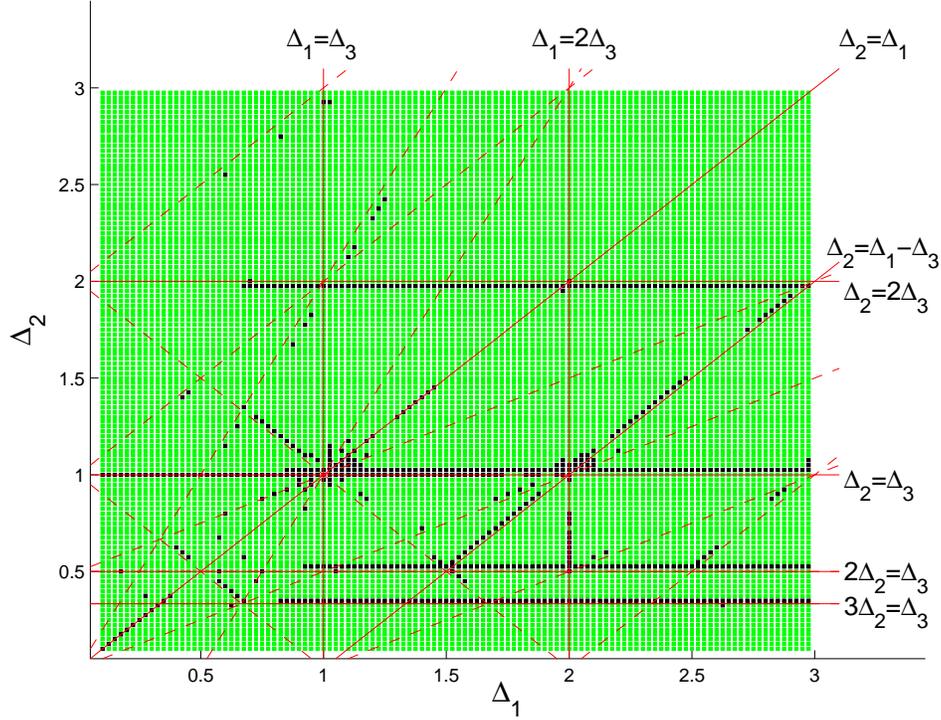}
\caption[]{$\Delta_1$-$\Delta_2$ plane for $N=7$
with $\Delta_3=1$ and $\varepsilon=0.3 (N/2) \min_i{\Delta_i}$. 
In the green regions the dynamics in $\mathcal{M}$ is transversally stable
({\em i.e.}~there is an attractor $\mathcal{A} \subseteq \mathcal{M}$). Inside the black regions
$\mathcal{M}$ is transversally unstable. (Red) lines indicate the loci
of the most important resonances among $\{\Delta_i\}$.}
\label{n7}
\end{center}
\end{figure}

For $N=7$ the system's dimension is too large to perform a detailed analysis around 
all relevant resonances.
But still the dimension is small enough to carry out intensive numerical simulations.
We fixed $\Delta_3=1$ and measured the largest TLE
at different values of $\Delta_1$ and $\Delta_2$. The initial condition in $\mathcal{M}$
was random (what should not be a problem if as we expect multistability is not common
at small $\varepsilon$).
In our simulations 
the coupling strength was $\varepsilon=0.3 (N/2) \min_i{\Delta_i}$. This value of $\varepsilon$ 
is small enough to ensure the systems is far from synchronization, 
and at the same time large enough to make convergence times
not exceedingly long for our computational resources. We may expect this value of $\varepsilon$ to capture
the fundamental phenomenology as $\varepsilon \to 0$. The result is presented in
Fig.~\ref{n7}, and shows that the dynamics on $\mathcal{M}$ is  transversally stable 
in almost all the $\Delta_1$-$\Delta_2$ plane,
except close to some resonances. These resonances should correspond to combinations
of $\Delta_i$'s not fulfilling the nonresonance condition of Th.~1 (but also one may not 
exclude finite-$\varepsilon$ effects).
The result is very much equivalent to the result for $N=5$, but with
unstable regions organized 
around resonances involving three instead of two frequencies.

\section{Even $N$}
In the symmetric Kuramoto model at small coupling the invariant torus $\mathcal{M}$ 
is almost always stable if the population size is odd. However, we show in this section that
if $N$ is an even number
$\mathcal{M}$ can be both stable and unstable in large regions of the
parameter space spanned by the natural frequencies.
We analyze the case $N=4$ in detail by means of the renormalization group (RG) method, 
and the case $N=6$ is studied using numerical calculations. Both cases share
features that should be common to any even number $N \ge 4$.

\subsection{$N=4$}

When $N=4$, Eq.~(\ref{kuramoto_model2}) with $\Delta_1=\Delta_3$ is written as
\begin{equation}
\left\{ \begin{array}{l}
\displaystyle \dot{\varphi }_1 = \Delta_1 + \frac{\varepsilon }{N} 
    \Bigl( -2 \sin \varphi _1 + \sin \varphi _2 \\
    \quad \quad  - \sin (\varphi _1 + \varphi _2) + \sin (\varphi _2 + \varphi _3)
                 - \sin (\varphi _1 + \varphi _2 + \varphi _3)\Bigr) ,\\
\displaystyle \dot{\varphi }_2 = \Delta_2 + \frac{\varepsilon }{N} 
    \Bigl( -2 \sin \varphi _2 + \sin \varphi _1 + \sin \varphi _3 
           - \sin (\varphi _1 + \varphi _2) - \sin (\varphi _2 + \varphi _3)\Bigr) , \\
\displaystyle \dot{\varphi }_3 = \Delta_1 + \frac{\varepsilon }{N} 
    \Bigl( -2 \sin \varphi _3 + \sin \varphi _2 \\
    \quad \quad  - \sin (\varphi _2 + \varphi _3) + \sin (\varphi _1 + \varphi _2)
                 - \sin (\varphi _1 + \varphi _2 + \varphi _3)\Bigr) .\\
\end{array} \right.
\label{N4}
\end{equation}
In this case, the invariant torus is given by the $2$-dimensional torus
$\mathcal{M} = \{ \varphi _1 = \varphi _3, \varphi_2 \}$.
Note that there will exist $n:m$ lockings on $\mathcal{M}$ like for $N=5$.

The linear ODE governing infinitesimal deviations off the torus ($\delta\varphi\equiv\varphi_1-\varphi_3$) is:
\begin{equation}
\dot{\delta\varphi}=-\frac{\varepsilon}{2}\left[\cos \varphi_1+\cos(\varphi_1+\varphi_2)\right]\delta\varphi ,
\label{dn4}
\end{equation}
and it determines the TLE.

In what follows, we suppose\footnote{If $\Delta_2=0$ there are two situations:
(i) If $\Delta_1=0 $, Eq.~(\ref{N4}) has a stable fixed point
$\varphi _i = 0,\,\, (i = 1,2,3)$ with $\lambda_\perp=-\varepsilon$.
(ii) If $\Delta _1 \neq 0$ there is a stable periodic orbit ($1:0$ locking solution).
Dynamics for $\varphi _2$ is well described  by averaging the second equation of Eq.~(\ref{N4})
with respect to $\varphi _{1}$ and $\varphi _{3}$:
$\dot{\varphi }_2 = -2\frac{\varepsilon }{N} \sin \varphi _2$.
It proves that $\varphi _2 = 0$ is stable. Substituting $\varphi _2 = 0$ into Eq.~(\ref{N4}),
we obtain the equation for $\varphi\equiv\varphi _1=\varphi _3$: 
$\dot{\varphi } = \Delta_1 - ({\varepsilon }/{N}) (2 \sin \varphi + \sin(2\varphi))$.
The transverse Lyapunov exponent is calculated using Eq.~(\ref{dn4}) in the same
way as that of $N=3$, obtaining $\lambda_\perp=-\varepsilon^3/(16 \Delta_1^2)$.}
$\Delta _2 \neq 0$. By dividing Eq.~(\ref{N4}) by $\Delta_2$,
we can assume $\Delta _2 = 1$ without loss of generality.\\

\textbf{Theorem 3.} There exists a non-negative number $\varepsilon _0 = \varepsilon _0(\Delta_1)$ 
such that if $0 < \varepsilon < \varepsilon _0$, the invariant torus $\mathcal{M}$ is asymptotically stable for $\Delta_1 > 1$
and unstable for $\Delta_1 <1$.
$\varepsilon _0(\Delta_1)$ tends to zero as $\Delta_1 \to 0, 1/3, 1/2, 1,2, 3, 4$.
The transverse Lyapunov exponent of $\mathcal{M}$ is of $O(\varepsilon ^5)$ if $0 < \varepsilon < \varepsilon _0$.
\\

\begin{figure}
\begin{center}
\includegraphics[]{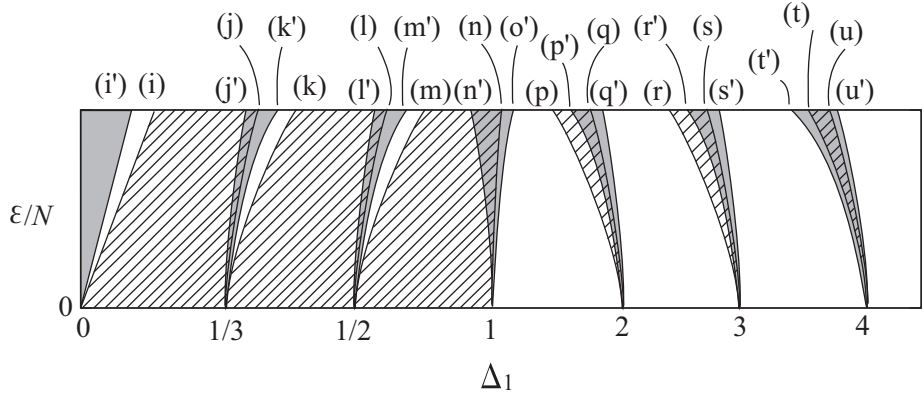}
\caption[]{A schematic view of the phase diagram of Eq.~(\ref{N4}) for small $\varepsilon $. 
There are no attractors $\mathcal{A} \subseteq \mathcal{M}$ in the hatched regions, and gray regions indicate the
Arnold tongues.} 
\label{tongue_n4}
\end{center}
\end{figure}

A sketch of the strategy for proving this theorem is given in Sec.~7.3,
while the detailed calculation is omitted.
A schematic view of the phase diagram of Eq.~(\ref{N4})
for small $\varepsilon $ is depicted in Fig.~\ref{tongue_n4},
in which there are no attractors $\mathcal{A} \subseteq \mathcal{M}$
in the hatched regions, and the $n:m$ locking solutions for 
$n:m = 0:1,\, 1:3,\, 1:2,\, 1:1,\, 2:1,\, 3:1$ and $4:1$ exist in the gray regions.
In particular in the regions that are gray but not hatched,
there are stable periodic orbits on $\mathcal{M}$.
Asymptotic expansions with respect to $\varepsilon /N$ of boundary curves (i) to (u) and (i') to (u')
in Fig.~\ref{tongue_n4} are shown in Appendix.

\begin{figure}
\begin{center}
\includegraphics[width=5.0in]{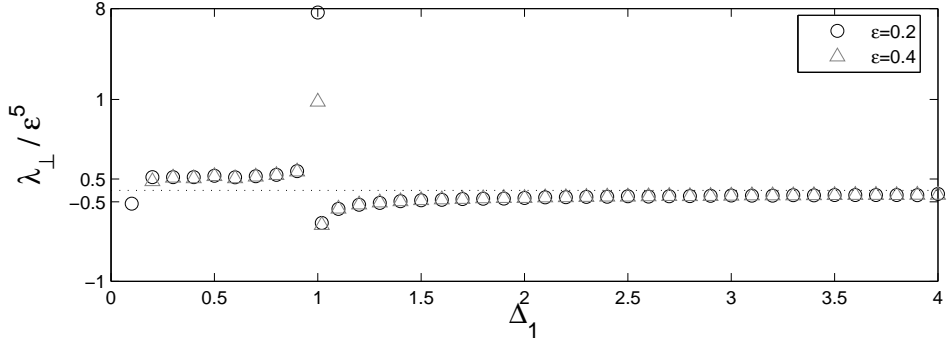}
\caption[]{Transverse Lyapunov exponent scaled by $\varepsilon^5$ 
as a function of $\Delta_1$ for two values of $\varepsilon$ ($N=4$, $\Delta_2=1$). The scale
of the y-axis is nonlinear to better discern the sign of the TLE.
There is a perfect overlap of both data sets ($\varepsilon=0.2$ and $0.4$), except at frequency lockings
$0:1$ and $1:1$. The stability change at $\Delta_1=1^+$ is very sharp but continuous.}
\label{num_n4}
\end{center}
\end{figure}

\begin{figure}
\begin{center}
\includegraphics[width=5.0in]{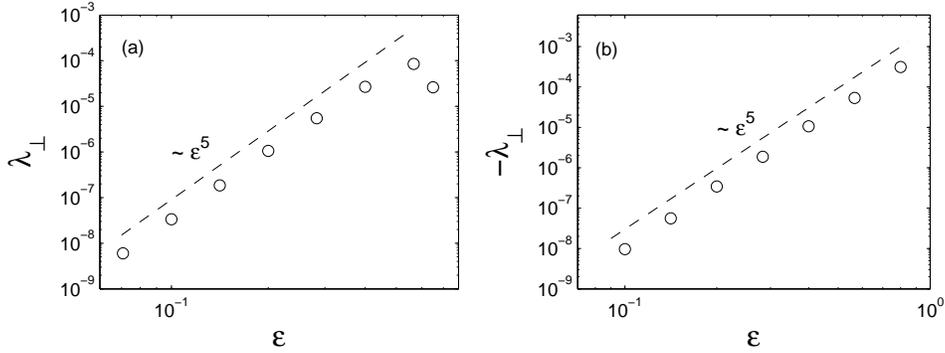}
\caption[]{TLEs as a function of $\varepsilon $ for $N=4, \Delta _2 = 1$ and (a) $\Delta _1 = (-1 + \sqrt{5})/2$ (the
inverse of the golden mean) (b) $\Delta _1 = (1 + \sqrt{5})/2$ (the golden mean). 
For the case (a) the invariant torus $\mathcal{M}$ is unstable while it is stable for (b)
as stated in Theorem 3.}
\label{n4_e5}
\end{center}
\end{figure}

We numerically calculated the TLE from Eq.~(\ref{dn4}), and 
Figs.~\ref{num_n4} and \ref{n4_e5} demonstrate that the TLE is $O(\varepsilon^5)$ as stated in Theorem 3.
$\mathcal{M}$ is mainly stable when $\Delta_1 > 1$. Like for $N=5$
there are ``switching tongues" in which the stability of $\mathcal{M}$ is different from the dominant stability
in its neighborhood.
We do not show them in Fig.~\ref{num_n4} not to overwhelm
the reader with details. We note nevertheless that the analytical expressions (see Appendix)
of boundary curves (i) to (u) have been corroborated by our numerical simulations.

\subsection{$N\geq 6$}

\begin{figure}
\begin{center}
\includegraphics[width=5.0in]{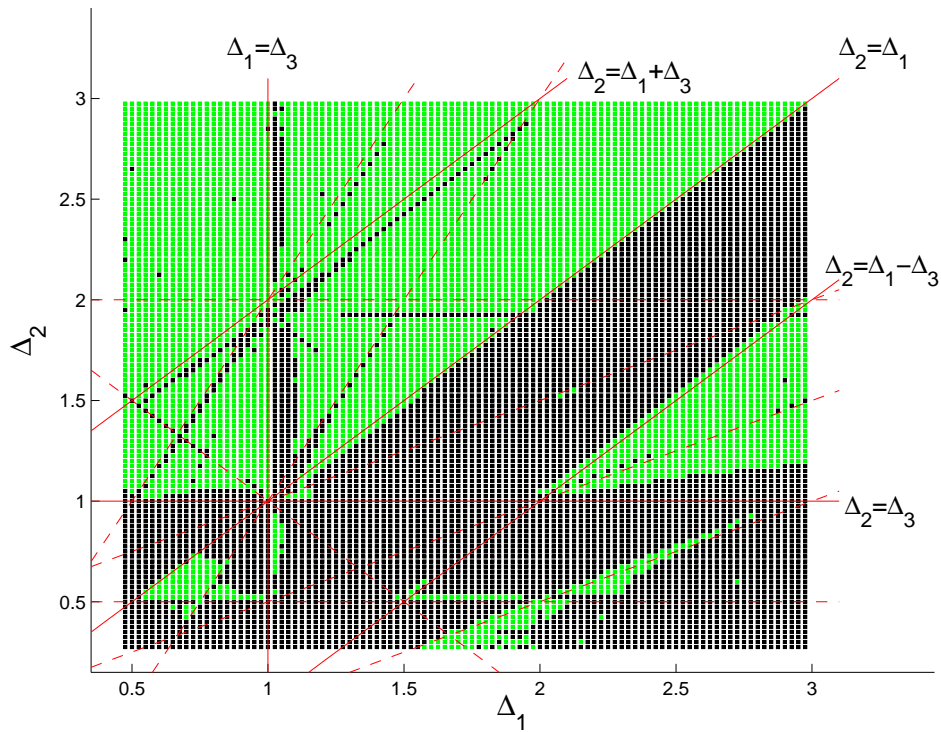}
\caption[]{$\Delta_1$-$\Delta_2$ plane for $N=6$
with $\Delta_3=1$ and $\varepsilon=0.3 (N/2) \min_i{\Delta_i}$. 
In the green regions the dynamics in $\mathcal{M}$ is transversally stable
({\em i.e.}~there is an attractor $\mathcal{A} \subseteq \mathcal{M}$). Inside the black regions
$\mathcal{M}$ is unstable.
(Red) lines indicate the loci
of the most important resonances among $\{\Delta_i\}$. In the regions with small $\Delta_1$ or $\Delta_2$
the convergence was too slow to guarantee accurate results (we have indications that typically
the TLEs are $O(\varepsilon^5)$ as for $N=4$).}
\label{n6}
\end{center}
\end{figure}

We resort to extensive numerical simulations to study the case $N=6$,
as we did already in Sec.~\ref{sec_n7} for $N=7$.
Figure \ref{n6} summarizes the result of our simulations.
There are regions with stable $\mathcal{M}$, and regions with unstable $\mathcal{M}$.
Some low order resonances give rise to the border among these regions (analogously to the $\Delta_1=\Delta_2$
border in the $N=4$ case). Other resonances give rise to thin strips where the 
stability switches (again in good analogy to the $N=4$ case).

If $\Delta_1$ is sufficiently large, oscillators $\theta _1$ and $\theta _6$ rotate so fast ($|\omega_{1}| \gg |\omega_{2,3}|$)
that their influence on the other oscillators averages out. In this regime, (in)stability of the invariant torus
is ruled by the four oscillators with the central frequencies. It may be perceived 
in Fig.~\ref{n6} that
for $\Delta_1 \gg \Delta_{2,3}$ there is a switch of stability at
$\Delta_2 \gtrapprox \Delta_3 =1$ in qualitative agreement 
with the $\Delta_1=\Delta_2+O(\varepsilon^2)$ border in the $N=4$ case (the transition
is not closer to $\Delta_2=1$ due to the finiteness of $\varepsilon$ and $\Delta _1$).

This simple argument based on the averaging method can be extended to an arbitrary (even) population size:
In the limit $\Delta_1\to\infty$, Eq.~(\ref{kuramoto_model2})
for $N = 2M+2$ becomes equivalent to Eq.~(\ref{kuramoto_model2}) for $N=2M$
because influence of $\varphi _1$ and $\varphi _{2M+1}$ on the other oscillators averages out.
Thus by induction, we conclude that the situation of $N=4$ is typical for general even $N$;
that is, regions of frequency space
where $\mathcal{M}$ is stable and regions where it is unstable 
coexist in parameter space, none of them disappearing as $\varepsilon\to 0$.
This property quite differs from the phase diagrams for odd $N$.

\section{Discussion}

One of the important results of our paper is the rather surprising fact that qualitative
properties of the Kuramoto model (with symmetrically allocated natural frequencies)
depend crucially on whether $N$ is odd or even.
We establish precise mathematical criteria for the stability of the invariant torus $\mathcal{M}$.
It is remarkable that in many cases this torus is asymptotically stable at arbitrarily small coupling.
For $N=4,5$, we have completely uncovered stability changes caused by resonances.

The renormalization group method has been successfully used in this paper.
In the literature, first- and second-order RG equations have been employed 
for constructing approximate solutions to weakly perturbed ODEs.
In this paper, we have used this technique in the Kuramoto model up to 
third and fifth order for odd $N$ and for $N=4$, respectively.
RG equations are quite helpful for studying the stability of invariant manifolds,
and they provide as well the orders of magnitude of the transverse Lyapunov exponents.

In the context of coupled oscillators, the phenomenon known as `phase chaos' consist 
in the appearance of a high-dimensional chaotic attractor \cite{topaj02,liu03,popovych05}
due to the interaction of phase variables (neutrally stable variables in the uncoupled limit).
For the Kuramoto model it has been reported  in \cite{popovych05} that 
(with a symmetric allocation of the natural frequencies)
when increasing the coupling from zero the Lyapunov exponents split from a degenerate set at zero
to a set of $[N/2]-1$ positive, $[N/2]-1$ negative, and 2 (3 if $N$ odd) zero Lyapunov exponents.
In this paper we show that the invariant torus $\mathcal{M}$ is often stable in the small coupling limit.
In contrast it is not proven yet that phase chaos indeed persists
in the $\varepsilon\to 0$ limit. Here we note that our Theorem 1 applies in a finite
range $0<\varepsilon<\varepsilon_0$ irrespective of how large $N$ is. 
Notice nevertheless that $\varepsilon_0 \to 0$ as $N\to\infty$.

Other implications of our paper refer to numerical simulations.
In this context our paper is particularly relevant because
the Kuramoto model has been usually considered together with a
symmetric frequency distribution of the natural frequencies, and in turn 
simulations, with a finite population size, are often carried out selecting frequencies
that reflect the inherent symmetry of the frequency density.
Our results also evidence the important role of resonances among the natural frequencies
for the stability of the invariant torus $\mathcal{M}$. This must serve as a warning about the risks of
using highly resonant frequencies when tackling generic properties of the system.
In particular, it has become popular to consider evenly spaced natural frequencies
($\Delta _i = \Delta _j$ for all $i,j$), which is probably the most resonant case.

We may conjecture that the results reported here for the Kuramoto model 
(with a sinusoidal coupling function) should also be observed in 
a family of odd coupling functions. In fact, in the $N=3$ case, see Eq.~(\ref{f}),
a family of functions shares stability and scaling of the transverse Lyapunov exponent.
Nonetheless in what concerns the unstable regions (tongue-shaped for $N=5$) and their scaling 
one may expect important differences depending on the coupling function $f$.
(We suspect this should be the case because of the similarity with the phase-locking 
regions, which depend on the harmonics of the interaction function~\cite{galkin94}.)

Finally, note that under a small enough symmetry-breaking perturbation 
$\mathcal{M}$ will get deformed into an invariant torus $\mathcal{M'}$ with the same stability.
Therefore our results may apply to situations where the symmetry is weakly broken.

\section{Outline of the proofs of theorems}

\subsection{Brief review of the RG method}

The renormalization group (RG) method is one of the singular perturbation methods for differential equations which
provides approximate solutions as well as approximate invariant manifolds and their stability.
Recently, it is shown that the RG method unifies and extends traditional singular perturbation methods, such as
the averaging method, the multi-time scale method, the normal forms theory and so on.
In this section, we give a brief review of the RG method following \cite{Chiba1,Chiba2}
to prove Theorems 1 in the next subsection.

Consider the system of differential equations on a compact manifold $M$ of the form
\begin{equation}
\frac{dx}{dt} = \dot{x} = \varepsilon g_1(t, x) + \varepsilon ^2 g_2(t,x) + \varepsilon ^3 g_3(t,x) + \cdots ,\,\, x\in M,
\label{RG1}
\end{equation}
where $\varepsilon \in \mathbb{R}$ is a small parameter.
For this system, we make the following assumption (A):

\textbf{(A)} \, The vector fields $g_i(t,x),\,\, i= 1,2,\cdots $ are $C^1$ with respect to time $t \in \mathbb{R}$ and 
$C^\infty$ with respect to $x\in M$.
Further, $g_i$ are almost periodic functions with respect to $t$ uniformly in $x \in M$,
the set of whose Fourier exponents has no accumulation points on $\mathbb{R}$.

In the case of the Kuramoto model (\ref{kuramoto_model}), $M$ is an $N$-dimensional torus.
Note that under the change of coordinates $\theta _j = x_j + \omega _j t$ and $\varphi _j = x_j + \Delta_j t$ 
systems (\ref{kuramoto_model}) and (\ref{kuramoto_model2}), respectively, are transformed into the form of Eq.~(\ref{RG1})
with $g_i=0$ for $i\ge2$, and satisfying the assumption (A).

Substitute $x = x_0 + \varepsilon x_1 + \varepsilon ^2 x_2 + \cdots $ into the right hand side of Eq.~(\ref{RG1}) and expand
it with respect to $\varepsilon $. We write the resultant as
\begin{equation}
\sum^{\infty}_{k=1}\varepsilon ^k g_k(t, x_0 + \varepsilon x_1 + \varepsilon ^2 x_2 + \cdots)
 = \sum^{\infty}_{k=1}\varepsilon ^k G_k(t, x_0, x_1, \cdots , x_{k-1}).
\end{equation}
For instance, $G_1, G_2$ and $G_3$ are given by
\begin{eqnarray}
 G_1(t,x_0) & =&  g_1(t, x_0), \\
 G_2(t,x_0,x_1)& =&  \frac{\partial g_1}{\partial x}(t,x_0)x_1 + g_2(t, x_0), \\
 G_3(t,x_0,x_1,x_2)& =&  \frac{1}{2}\frac{\partial ^2g_1}{\partial x^2}(t,x_0)x_1^2
 + \frac{\partial g_1}{\partial x}(t, x_0)x_2 + \frac{\partial g_2}{\partial x}(t,x_0)x_1 + g_3(t, x_0),
\end{eqnarray}
respectively.
With these $G_i$'s, we define the $C^\infty$ maps $R_i, u^{(i)}_t : M \to M$ to be
\begin{eqnarray}
& & R_1(y) = \lim_{t\to \infty}\frac{1}{t} \int ^t\! G_1(s,y)ds, \\
& & u^{(1)}_t(y) = \int^t \! \left( G_1(s,y) - R_1(y) \right) ds,
\end{eqnarray}
and 
\begin{eqnarray}
& & R_i(y)  
= \lim_{t\to \infty}\frac{1}{t} \int ^t\! \Bigl(
G_i(s, y, u^{(1)}_s(y),
 \cdots, u^{(i-1)}_s(y))- \sum^{i-1}_{k=1}\frac{\partial u_s^{(k)}}{\partial y}(y)R_{i-k}(y) \Bigr) ds , \\
& & u^{(i)}_t(y)= \int^t \! \Bigl( G_i(s, y,u^{(1)}_s(y)
, \cdots , u^{(i-1)}_s(y)) - \sum^{i-1}_{k=1}\frac{\partial u_s^{(k)}}{\partial y}(y) R_{i-k}(y) - R_i(y) \Bigr) ds,
\end{eqnarray}
for $i=2,3,\cdots$, respectively, where $\int^t$ denotes the indefinite integral, whose integral constants are 
fixed arbitrarily.
We can prove that $R_i$ are well-defined ({\em i.e.}~the limits exist) and $u^{(i)}_t$ are bounded in $t\in \mathbb{R}$.
Along with $R_i$ and $u^{(i)}_t$, we define the 
\textit{$m$-th order RG equation} for Eq.~(\ref{RG1}) to be
\begin{equation}
\dot{y} = \varepsilon R_1(y) + \varepsilon ^2R_2(y) + \cdots  + \varepsilon ^mR_m(y), \,\, y\in M,
\label{RG2}
\end{equation}
and the \textit{$m$-th order RG transformation} $\alpha ^{(m)}_t$ to be
\begin{equation}
\alpha ^{(m)}_t (y) = y + \varepsilon u^{(1)}_t(y) + \cdots + \varepsilon ^m u^{(m)}_t(y), \,\, y\in M.
\label{RG2b}
\end{equation}

Roughly speaking, we can show that the $m$-th order RG transformation $x = \alpha ^{(m)}_t (y)$
brings the system (\ref{RG1}) into the system of the form $\dot{y} = \varepsilon R_1(y) + \cdots  + \varepsilon ^mR_m(y) + \varepsilon ^{m+1}S(t,x, \varepsilon )$,
where $S$ is bounded in $t \in \mathbb{R}$.
It means that the $m$-th order RG equation is $\varepsilon ^{m+1}$-close to the original system (\ref{RG1})
and thus it is useful to construct the flow of (\ref{RG1}) approximately.
Since the RG equation is an autonomous system while the original system (\ref{RG1}) is not,
to analyze the RG equation is easier than that of the original system.
The next theorem is one of the fundamental theorems of the RG method.

\textbf{Theorem A \cite{Chiba1,Chiba2}.} Suppose that $R_1(y) = \cdots = R_{k-1}(y) = 0$ and
$\varepsilon ^k R_k(y)$ is the first non-zero term in the RG equation.
If the vector field $R_k(y)$ has a boundaryless compact normally hyperbolic invariant manifold $\mathcal{N}$,
then for sufficiently small $\varepsilon >0$,
Eq.~(\ref{RG1}) has an invariant manifold $\mathcal{N}_\varepsilon $, which is diffeomorphic to $\mathcal{N}$.
In particular, stability of $\mathcal{N}_\varepsilon $ coincides with that of $\mathcal{N}$.

This theorem is used to investigate the stability of the invariant torus $\mathcal{M}$ and the 
$n:m$ locking solutions of the Kuramoto model.

\subsection{Proof of Theorem 1}

In this section, we give the proof of Theorem 1.

\textbf{Proof of Theorem 1.} Suppose that $N = 2M-1$ is an odd number and $\omega _i$'s are allocated 
symmetrically as was assumed.
Put 
$\theta  _i = x_i + \omega _i t $
and rewrite Eq.~(\ref{kuramoto_model}) of the form of Eq.~(\ref{RG1}).
If the natural frequencies satisfy the nonresonance condition (\ref{NRcondition}), its third-order RG equation
is given by
\begin{equation}
\left\{ \begin{array}{l}
\displaystyle \dot{y }_M
   = -\frac{16 \varepsilon ^3}{N^3}\sum_{k\neq M} \frac{1}{\omega _k^2} \sin (2y _M - y _k - y _{2M-k}),  \\
\displaystyle \dot{y }_i
   = \frac{8\varepsilon ^2}{N^2}\left( 2\sum_{k\neq i} \frac{1}{\omega _i - \omega _k}
               - \frac{1}{\omega _i} \cos (y _i - 2y _M + y _{2M-i})\right)  \\
\displaystyle \quad + \frac{16\varepsilon ^3}{N^3}\Biggl( \sum_{k \neq i, 2M-i} 
                      \frac{1}{\omega _i^2 - \omega _k^2} \sin (y _i - y _k - y _{2M-k} + y _{2M-i}) \\
\displaystyle \quad -2\sum_{k\neq i,2M-i}\frac{1}{\omega _i (\omega _i - \omega _k)}
                      \sin (y _i - y _k - y _{2M-k} + y _{2M-i}) \\
\displaystyle \quad + 2\sum_{k \neq i, M} \frac{1}{\omega _k (\omega _i - \omega _k)}
                     \sin (2y _M - y _k - y _{2M-k}) \\
\displaystyle \quad - \sum_{k\neq M,i,2M-i} \frac{1}{\omega _k (\omega _i + \omega _k)}
                       \sin (y _i - y _k - y _{2M-k} + y _{2M-i}) \\
\displaystyle \quad - 2 \sum_{k\neq M, 2M-i} \frac{1}{\omega _i (\omega _i + \omega _k)}
                        \sin (y _i - 2y _M + y _{2M-i})\Biggr),\,\,\,\, (i \neq M).
\end{array} \right.
\label{RG3}
\end{equation}
Note that the first order term vanishes and the expansion begins with the second order term.
Since the invariant torus $\mathcal{M}$ corresponds to the solution $y _i + y_{2M-i} = c$ (constant),
we put $\phi_i = y _i + y _{2M-i}$ and $\phi_M = 2y _M$.
Then we obtain the system of $\phi_i$ :
\begin{equation}
\left\{ \begin{array}{l}
\displaystyle \dot{\phi}_M
    = -\frac{64 \varepsilon ^3}{N^3}  \sum^{M-1}_{k=1} \frac{1}{\omega _k^2} \sin (\phi_M - \phi_k), \\
\displaystyle \dot{\phi}_i
    = \frac{32\varepsilon ^3}{N^3} \Biggl( -\frac{1}{\omega _i^2} \sin (\phi_i - \phi_M)
       - 4 \sum^{M-1}_{k\neq i} \frac{1}{\omega _i^2 - \omega _k^2} \sin (\phi_i - \phi_M) \\
\displaystyle \quad \quad \quad \quad  + 4 \sum^{M-1}_{k\neq i} \frac{1}{\omega _i^2 - \omega _k^2} 
              \sin (\phi_M - \phi _k)\Biggr),\,\,\, (i=1,\cdots ,M-1).
\end{array} \right.
\label{RG4}
\end{equation}
Now that the second order term vanishes and Theorem A for $k=3$ is applicable to this system.
We can prove 
that the eigenvalues of the Jacobian matrix of the r.h.s.~of (\ref{RG4}) at the fixed point 
$\phi_i = c\,\, (i=1, \cdots  ,M)$ have
negative real parts, except a zero eigenvalue that results from the rotation invariance of Eq.~(\ref{kuramoto_model})
(or the degree of freedom of the constant $c$).
A proof of this fact is outlined as follows:

Let $\mathcal{J}$ be the Jacobian matrix of the r.h.s.~of (\ref{RG4}) at the fixed point 
$\phi_i = c\,\, (i=1, \cdots  ,M)$. By using the cofactor expansion,
it is easy to show that the characteristic polynomial of $\mathcal{J}$
is calculated as
\begin{eqnarray}
\det\, (\lambda I - \frac{N^3}{32 \varepsilon ^3}\mathcal{J})
  = \lambda \cdot \det\, (\lambda I + \mathcal{A}_{M-1}),
\label{revised1}
\end{eqnarray}
where the matrix $\mathcal{A}_{M-1}$ is given as
\begin{eqnarray}
\mathcal{A}_{M-1} = \left(
\begin{array}{@{\,}cccc@{\,}}
\displaystyle \frac{3}{\omega _1^2} + \sum^{M-1}_{k\neq 1} \frac{4}{\omega ^2_1 - \omega ^2_k}
  & \displaystyle \frac{4}{\omega _1^2 - \omega _2^2} + \frac{2}{\omega_2^2} 
    & \cdots & \displaystyle \frac{4}{\omega _1^2 - \omega _{M-1}^2} + \frac{2}{\omega_{M-1}^2} \\
\displaystyle \frac{4}{\omega _2^2 - \omega _1^2} + \frac{2}{\omega_1^2} 
  & \displaystyle  \frac{3}{\omega _2^2} + \sum^{M-1}_{k\neq 2} \frac{4}{\omega ^2_2 - \omega ^2_k}
    & \cdots & \displaystyle \frac{4}{\omega _2^2 - \omega _{M-1}^2} + \frac{2}{\omega_{M-1}^2} \\
\vdots & \vdots  & \ddots & \cdots \\
\displaystyle \frac{4}{\omega _{M-1}^2 - \omega _{1}^2} + \frac{2}{\omega_{1}^2} 
  & \displaystyle \frac{4}{\omega _{M-1}^2 - \omega _{2}^2} + \frac{2}{\omega_{2}^2}
    & \cdots 
   & \displaystyle \frac{3}{\omega _{M-1}^2} + \sum^{M-1}_{k\neq M-1} \frac{4}{\omega ^2_{M-1} - \omega ^2_k} \\
\end{array}
\right).
\label{revised2}
\end{eqnarray}
Eq.~(\ref{revised1}) shows that $\mathcal{J}$ has a zero eigenvalue $\lambda =0$.
Now it is sufficient to prove that all eigenvalues of $\mathcal{A}_{M-1}$ have positive real parts.
To prove it, let 
\begin{equation}
\lambda ^{M-1} + f^{(M-2)}_{M-1} \lambda ^{M-2} + f^{(M-3)}_{M-1} \lambda ^{M-3}
 + \cdots + f^{(1)}_{M-1} \lambda + f^{(0)}_{M-1}=0
\end{equation}
be the characteristic equation $\det (\lambda I - \mathcal{A}_{M-1}) = 0$ of $\mathcal{A}_{M-1}$. 
We show the inequalities $f^{(M-2)}_{M-1}, f^{(M-4)}_{M-1}\cdots <0$ and $f^{(M-3)}_{M-1}, f^{(M-5)}_{M-1} \cdots  >0$
by induction on $M$. Since $f^{(i)}_{M-1}$ is invariant under the permutation of $\alpha _1 ,\cdots , \alpha _{M-1}$,
we can show that $f^{(i)}_{M-1}$ is of the form
\begin{equation}
f^{(i)}_{M-1} = b^{(i)}_{M-1}\frac{\sum_{i_1< \cdots < i_k} \alpha ^2_{i_1} \alpha ^2_{i_2} \cdots \alpha ^2_{i_k}}
{\alpha ^2_1 \alpha ^2_2 \cdots \alpha ^2_{M-1}},\,\, b^i_{M-1} \in \mathbb{R}.
\end{equation}
Since $\mathcal{A}_{M-1} \to \left(
\begin{array}{@{\,}cc@{\,}}
\mathcal{A}_{M-2} & 0 \\
\textstyle{*} & 0
\end{array}
\right)$ as $\alpha _{M-1} \to 0$, $f^{(i)}_{M-1} \to f^{(i-1)}_{M-2}$ as $\alpha _{M-1} \to 0$.
Now induction on $M$ proves the desired inequalities.

Thus, the solution $\phi_i = y _i + y _{2M-i}=c \,\, (i=1, \cdots ,M)$ of the RG equation is asymptotically stable and
this proves that the invariant torus $\mathcal{M}$ is asymptotically stable for small $\varepsilon >0$.
Note that the degree of freedom of $c$ does not appear in the $\varphi _j$ coordinates [Eq.~(\ref{kuramoto_model2})].
\hfill $\blacksquare$

If $N=3 (M=2)$, the nonresonance condition (\ref{NRcondition}) is reduced to $\Delta_1 \neq 0$
and Theorem 1 recovers the results obtained in Section 4.1.

\subsection{Sketch of the proofs of Theorems 2 and 3}

Theorems 2 and 3 are also proved by using the RG method though we need much harder analysis
to obtain asymptotic expansions of the boundary lines in Figs.~\ref{tongue_n5} and \ref{tongue_n4}.
In this section, we offer the strategy to prove Theorems and to derive the asymptotic expansions,
which is also valid for any $N$.

In what follows, we assume $\Delta_2 = 1$ in Eqs.~(\ref{N5}), (\ref{N4}) as was mentioned.
Our strategy to prove Theorems 2 and 3, and to obtain boundary lines, 
is summarized as follows:

\begin{enumerate}
\item[(i)]
Put $\varphi _i = x_i + \Delta_i t$ and rewrite Eqs.~(\ref{N5}) and (\ref{N4})
into the form of Eq.~(\ref{RG1}).

\item[(ii)]
Derive the RG equations up to third-order for $N=5$ and to fifth-order for $N=4$.
The forms of RG equations depend on $\Delta_1$.
Find the set of values $\Delta_1$, which gives the nonresonance condition, at which
RG equations take different forms from the others.
We find that the nonresonance conditions are given by $\Delta_1 \neq 0, 1/2, 1, 2, 3, 4$ for $N=5$
and $\Delta_1 \neq 0, 1/3, 1/2, 1, 2, 3, 4$ for $N=4$.

\item[(iii)]
Investigate the stability of the invariant torus for the RG equation satisfying
the nonresonance condition as was done in Sec.~7.2.
In this step, the proof of Theorems 2 and 3 ends.

\item[(iv)]
To find the Arnold tongues in Figs.~\ref{tongue_n5} and \ref{tongue_n4}, let $c_0$
be a resonance value obtained in step (ii).
Put 
\begin{equation}
\Delta _1 = c_0 + c_1 \varepsilon /N + c_2 \varepsilon ^2 /N^2 + \cdots
\label{sec7.3}
\end{equation}
in Eqs.~(\ref{N5}) and (\ref{N4}), and derive the RG equations.

\item[(v)]
Investigate the stability of the invariant torus for the resultant RG equations
and find values $c_1, c_2, \cdots $ at which the stability changes.
Then, Eq.~(\ref{sec7.3}) gives an asymptotic expansion of a boundary line of the Arnold tongue
emerging from $\Delta_1 = c_0 $ in the phase diagram.

\end{enumerate}

In the next subsection, we calculate asymptotic expansions of boundary lines
(a), (b), (a') and (b') in  Fig.~\ref{tongue_n5} to confirm this strategy for
$N=5$ and $\Delta_1$ near $1/2$.
Other expansions of boundary lines (c) to (u) and (c') to (u') in Figs.~\ref{tongue_n5}
and \ref{tongue_n4}
are obtained analogously, and their derivation is omitted; 
the results are given in the Appendix.

\subsection{Phase diagram near $\Delta_1=1/2$ for $N=5$}

In this section, 
we derive asymptotic expansions of lines (a) and (b), 
as well as asymptotic expansions of lines (a') and (b'), 
which are boundaries of the $1:2$ Arnold tongue for $N=5$.
We also show that there are many disjoint unstable regions of $\mathcal{M}$ emerging from 
$\Delta_1/\Delta_2 = 1/2$ as is shown in Fig.~\ref{tongue_n5}.

To investigate the phase diagram of $N=5$ near $\Delta_1/\Delta_2 = 1/2$, put
$\Delta _2 = 1,\, \Delta _1 = 1/2 + c_1 \varepsilon /N + c_2 \varepsilon ^2 /N^2 + \cdots $ 
and put $\varphi _1 = x_1 + t/2,\, \varphi _2 = x_2 + t,\, \varphi _3 = x_3 + t$ and 
$\varphi _4 = x_4 + t/2$ in Eq.~(\ref{N5}).
Then Eq.~(\ref{N5}) takes the form of Eq.~(\ref{RG1}) and the RG method is applicable.
The third-order RG equation for the system is given by
\begin{eqnarray}
& & \frac{d}{dt}\left(
\begin{array}{@{\,}c@{\,}}
y_1 \\
y_2 \\
y_3 \\
y_4
\end{array}
\right) = \frac{\varepsilon }{N}\left(
\begin{array}{@{\,}c@{\,}}
 c_1 \\
0 \\
0 \\
c_1
\end{array}
\right) \nonumber \\
& & + \frac{\varepsilon ^2}{N^2} \left(
\begin{array}{@{\,}c@{\,}}
\displaystyle -\frac{7}{2} + c_2 - \frac{1}{2} \cos (y_2-y_3) + \frac{1}{3} \cos (y_1+y_2-y_3-y_4) \\[0.1cm]
\displaystyle \frac{1}{10} + \frac{1}{2}\cos (y_2-y_3) \\[0.1cm]
\displaystyle \frac{1}{10} + \frac{1}{2}\cos (y_2-y_3) \\[0.1cm]
\displaystyle -\frac{7}{2} + c_2 - \frac{1}{2} \cos (y_2-y_3) + \frac{1}{3} \cos (y_1+y_2-y_3-y_4)
\end{array}
\right) \nonumber \\
& & \!\!\!\!\!\! \!\!\!\!\!\! 
+ \frac{\varepsilon ^3}{N^3}\left(
\begin{array}{@{\,}c@{\,}}
\displaystyle \frac{26 c_1}{3} + c_3 - \frac{2c_1}{9}\cos (y_1+y_2-y_3-y_4) - \frac{1}{2} \sin (2 y_1-y_2)
 \quad \quad \quad \quad  \quad \quad \quad \quad \\[0.1cm]
\displaystyle  \quad \quad \quad \quad
     -\frac{27}{20}\sin (y_2-y_3) - \frac{1}{12} \sin (y_2-2 y_4) - \frac{77}{45} \sin (y_1+y_2-y_3-y_4) \\[0.1cm]
\displaystyle -\frac{96 c_1}{25} - 2\sin (2 y_1-y_2) + \frac{1}{12} \sin (2 y_1-y_3) + \frac{1}{20} \sin (y_2-y_3)
 \quad \quad \quad \quad \quad \quad \quad \quad \\[0.1cm]
\displaystyle \quad \quad \quad \quad
      + \frac{1}{6}\sin (y_2-2 y_4) - \frac{3}{2} \sin (y_3-2 y_4) + \frac{26}{45}\sin (y_1+y_2-y_3-y_4) \\[0.1cm]
\displaystyle -\frac{96 c_1}{25} + \frac{3}{2} \sin (2 y_1-y_2) - \frac{1}{6}\sin (2 y_1-y_3) - \frac{1}{20} \sin (y_2-y_3)
 \quad \quad \quad \quad \quad \quad \quad \quad \\[0.1cm]
\displaystyle   \quad \quad \quad \quad
      - \frac{1}{12} \sin (y_2-2 y_4) + 2\sin (y_3-2 y_4) - \frac{26}{45} \sin (y_1+y_2-y_3-y_4) \\[0.1cm]
\displaystyle \frac{26 c_1}{3} + c_3 - \frac{2c_1}{9}\cos (y_1+y_2-y_3-y_4) + \frac{1}{12} \sin (2 y_1-y_3)
 \quad \quad \quad \quad \quad \quad \quad \quad \\[0.1cm]
\displaystyle \quad \quad \quad \quad
      + \frac{27}{20}\sin (y_2-y_3) + \frac{1}{2} \sin (y_3-2 y_4) + \frac{77}{45} \sin (y_1+y_2-y_3-y_4)
\end{array}
\right) .
\label{proof1}
\end{eqnarray}
This system has the solution $y_1 = y_4,\, y_2 = y_3$, which corresponds to the invariant torus $\mathcal{M}$.
The linearized equation for Eq.~(\ref{proof1}) along the solution $y_1 = y_4,\, y_2 = y_3$ is given as
\begin{eqnarray}
\frac{d}{dt}\left(
\begin{array}{@{\,}c@{\,}}
\delta y_1  \\
\delta y_2 
\end{array}
\right) =  \frac{\varepsilon ^3}{N^3} \left(
\begin{array}{ll}
\displaystyle  -\frac{7}{90} (44+15 \cos y) & \displaystyle -\frac{551}{90}+\frac{5 \cos y}{12} \\
\displaystyle  \frac{52}{45} - \frac{13 \cos y}{2} & \displaystyle \frac{113}{90} + \frac{15 \cos y}{4}
\end{array}
\right) \left(
\begin{array}{@{\,}c@{\,}}
\delta y_1 \\
\delta y_2 
\end{array}
\right),
\label{proof2}
\end{eqnarray}
where $\delta y_1 = y_1 - y_4, \delta y_2 = y_2 - y_3$,
and where $y = y(t)$ is a solution of the equation
\begin{equation}
\frac{dy}{dt} = \frac{2\varepsilon }{N}c_1 + \frac{\varepsilon ^2}{N^2} \left( 2c_2 - \frac{119}{15} \right) 
 + \frac{\varepsilon ^3}{N^3} \left( \frac{4664}{225}c_1 + 2 c_3 - \frac{1}{4} \sin y \right),
\label{proof3}
\end{equation} 
which is obtained by putting $2y_1 - y_2 = y,\, y_4 = y_1,\, y_3 = y_2$ in Eq.~(\ref{proof1}).
The existence of $1:2$ locking corresponds to the existence of a stable fixed point of $y$, and
the stability of the invariant torus $\mathcal{M}$ is 
maps to the stability of the trivial solution
$\delta y_1 = \delta y_2 = 0$ in Eq.~(\ref{proof2}).
Stability depends on the coefficients $c_i$ in Eq.~(\ref{proof3}).

\begin{enumerate}
\item[(i)]
When $c_1 \neq 0$, then we can apply the averaging method to Eqs.~(\ref{proof2}, \ref{proof3}).
Averaged with respect to $y$, Eq.~(\ref{proof2}) is rewritten as
\begin{equation}
\frac{d}{dt}\left(
\begin{array}{@{\,}c@{\,}}
\delta y_1  \\
\delta y_2 
\end{array}
\right) =  \frac{\varepsilon ^3}{N^3} \left(
\begin{array}{ll}
-154/45 \, \, & -551/90 \\
52/45 & 113/90
\end{array}
\right) \left(
\begin{array}{@{\,}c@{\,}}
\delta y_1 \\
\delta y_2 
\end{array}
\right).
\label{proof_s}
\end{equation}
It is easy to verify that the trivial solution of this system is stable because the eigenvalues of the matrix
in the right hand side are given by $(-13 \pm \sqrt{231}i)/12$.
It proves that the invariant torus $\mathcal{M}$ is stable if $c_1 \neq 0$.

\item[(ii)]
When $c_1 = 0$ and $c_2 \neq 119/30$, we can apply the averaging again to obtain Eq.~(\ref{proof_s})
what proves that $\mathcal{M}$ is stable in the same way as (i).

\item[(iii)]
When $c_1 = 0$ and $c_2 = 119/30$, Eq.~(\ref{proof3}) becomes
\begin{equation}
\frac{dy}{dt} = \frac{2\varepsilon ^3}{N^3} \left( c_3 - \frac{1}{8} \sin y \right).
\label{proof4}
\end{equation} 

\begin{enumerate}
\item[(iii-a)]
If $|c_3| < 1/8$, the above equation has a stable fixed point $y = y_*$ such that
$\sin y_* = 8c_3,\, \cos y_* = (1-64c_3^2)^{1/2}$.
It corresponds to the $1:2$ locking solution because $y = 2y_1 - y_2$. The disappearance of the fixed point 
at $|c_3|=1/8$ marks the
boundaries of the Arnold tongue: asymptotic expansions (a') and (b') in Appendix.
It is easy to investigate the stability of the trivial solution
of the linearized equation (\ref{proof2}) with constant coefficients.
Indeed, we can show that the trivial solution is unstable if and only if 
\begin{equation}
|c_3| < \frac{1}{4}\left( \frac{8681209}{17391218+5894\sqrt{8738809}}\right)^{1/2}=0.124838\ldots
\end{equation}
And this proves that the invariant torus $\mathcal{M}$ is unstable in the region surrounded by the lines (a) and (b)
given in Appendix.

\item[(iii-b)]
If $|c_3| > 1/8$,  the linearized equation (\ref{proof2}) is a linear system with a time periodic coefficient.
It is well known that stability of a trivial solution of such a system is determined by the 
Floquet exponents although we can not calculate them analytically in general.
We examine the stability of the trivial solution of Eq.~(\ref{proof2}) by calculating the Lyapunov exponents
numerically.

\begin{figure}
\begin{center}
\includegraphics[width=5.0in]{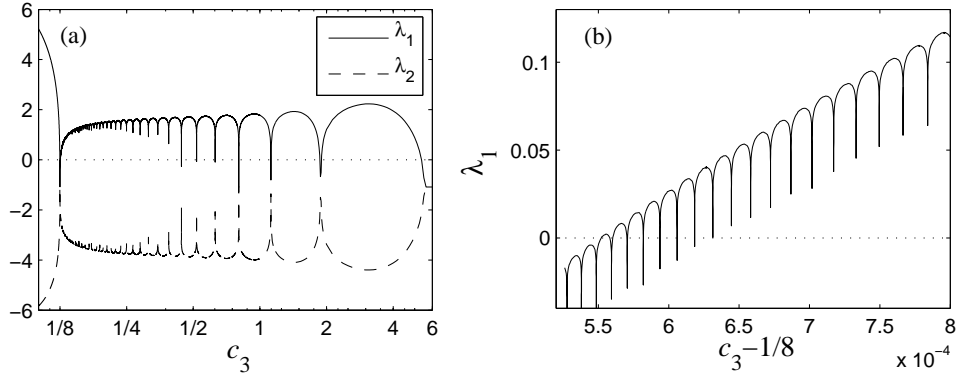}
\caption[]{(a) Lyapunov exponents ($\lambda_1\ge\lambda_2$) of system (\ref{proof2})
forced by (\ref{proof4}) ($\varepsilon/N=1$ is arbitrarily adopted).
In the interval  $0.125552\ldots < c_3 < 5.3695\ldots$, $\lambda_1$ is mostly positive except
at hundreds of narrow windows where it becomes negative (to coincide with $\lambda_2$).
(b) In this panel we see how $\lambda_1$ presents dips that accumulate at $c_3=1/8$;
the dips are so narrow that the sampling is not able to resolve the intervals where 
$\lambda_1=\lambda_2$.}
\label{app:numerics}
\end{center}
\end{figure}

Figure \ref{app:numerics} shows that there are many disjoint intervals of $c_3$ on which the trivial solution of Eq.~(\ref{proof2}) is unstable. 
It proves that there are many disjoint unstable regions of the invariant torus $\mathcal{M}$
emerging from $\Delta _1 = 1/2$. These unstable regions are inside the region limited by 
$\Delta_1 = \frac{1}{2} + \frac{119}{30}\frac{\varepsilon^2 }{N^2}
 + 0.125552 \frac{\varepsilon^3 }{N^3} + O(\varepsilon ^4) $
and
$ \Delta_1 = \frac{1}{2} + \frac{119}{30}\frac{\varepsilon^2 }{N^2}
 + 5.3695 \frac{\varepsilon^3 }{N^3} + O(\varepsilon ^4)$;
and inside a twin region with opposite signs in the cubic terms (dotted regions in Fig.~\ref{tongue_n5}).

\end{enumerate}

\end{enumerate}

\section*{Acknowledgments}

D.P.~acknowledges supports by CSIC under the Junta de Ampliaci\'on de
Estudios Programme (JAE-Doc), and by Ministerio de Educaci\'on
y Ciencia (Spain) under project No.~FIS2006-12253-C06-04.

\bibliographystyle{elsart-num}

\section*{Appendix : 
Asymptotic expansions of boundary curves in Fig.~\ref{tongue_n5} ($N=5$) and Fig.~\ref{tongue_n4} ($N=4$)}

Asymptotic expansions in $\varepsilon$ of lines (a) to (h) in Fig.~\ref{tongue_n5},
which divide stable regions and unstable regions of the invariant torus $\mathcal{M}$ for $N=5$, are
\begin{eqnarray*}
&(\rm{a,b})&  \Delta_1 = \frac{1}{2} + \frac{119}{30}\frac{\varepsilon^2 }{N^2}
 \mp \frac{1}{4}\left( \frac{8681209}{17391218+5894\sqrt{8738809}}\right)^{1/2}
\frac{\varepsilon^3 }{N^3} + O(\varepsilon ^4) , \\
&(\rm{c})&  \Delta_1 = 1 + \frac{2}{3}\frac{\varepsilon^2 }{N^2} - \frac{689009}{362880}\frac{\varepsilon^4 }{N^4}
 + O(\varepsilon ^5), \\
&(\rm{d})&  \Delta_1 = 1 + \frac{5}{3}\frac{\varepsilon^2 }{N^2}
 + \left( \frac{1027631}{241920} + k_1\right) \frac{\varepsilon^4 }{N^4} + O(\varepsilon ^5), \\
&(\rm{e})&  \Delta_1 = 2 - \frac{5}{3}\frac{\varepsilon^2 }{N^2} - \frac{139753}{241920}\frac{\varepsilon^4 }{N^4}
 + O(\varepsilon ^5), \\
&(\rm{f})&  \Delta_1 = 2 - \frac{2}{3}\frac{\varepsilon^2 }{N^2}+
\left( -\frac{136655}{48384} + k_2\right)\frac{\varepsilon^4 }{N^4} + O(\varepsilon ^5), \\
&(\rm{g,h})&  \Delta_1 = 3 -\frac{161}{60}\frac{\varepsilon^2 }{N^2}
 \mp \left( \frac{191\sqrt{18481} - 24281}{28800}\right)^{1/2}\frac{\varepsilon^3 }{N^3}+ O(\varepsilon ^4),
\end{eqnarray*}
where $k_1$ and $k_2$ are some positive constant which are not obtained analytically.

Asymptotic expansions of boundaries (a') to (h') of the $n:m$ Arnold tongues in Fig.~\ref{tongue_n5}
are
\begin{eqnarray*}
&(\rm{a',b'})&  \Delta_1 = \frac{1}{2} + \frac{119}{30}\frac{\varepsilon^2 }{N^2}
 \mp \frac{1}{8}\frac{\varepsilon^3 }{N^3} + O(\varepsilon ^4), \\
&(\rm{c'})&  \Delta_1 = 1 + \frac{2}{3}\frac{\varepsilon^2 }{N^2}
 - \frac{684751}{241920}\frac{\varepsilon^4 }{N^4} + O(\varepsilon ^5), \\
&(\rm{d'})&  \Delta_1 = 1 + \frac{5}{3}\frac{\varepsilon^2 }{N^2}
 + \frac{1027631}{241920}\frac{\varepsilon^4 }{N^4} + O(\varepsilon ^5), \\
&(\rm{e'})&  \Delta_1 = 2 - \frac{5}{3}\frac{\varepsilon^2 }{N^2}
 - \frac{291373}{241920}\frac{\varepsilon^4 }{N^4} + O(\varepsilon ^5), \\
&(\rm{f'})&  \Delta_1 = 2 - \frac{2}{3}\frac{\varepsilon^2 }{N^2}
 - \frac{136655}{48384}\frac{\varepsilon^4 }{N^4} + O(\varepsilon ^5), \\
&(\rm{g',h'})&  \Delta_1 = 3 - \frac{161}{60}\frac{\varepsilon^2 }{N^2}
 \mp \frac{1}{3}\frac{\varepsilon^3 }{N^3} + O(\varepsilon ^4).
\end{eqnarray*}

Asymptotic expansions of the lines (i) to (u) in Fig.~\ref{tongue_n4}, which divide stable regions and unstable regions
of the invariant torus $\mathcal{M}$ for $N=4$, are
\begin{eqnarray*}
&(\rm{i})&  \Delta_1 = \frac{58}{\sqrt{805}}\frac{\varepsilon }{N} + O(\varepsilon ^2) , \\
&(\rm{j})&  \Delta_1 = \frac{1}{3} + \frac{193}{30}\frac{\varepsilon^2 }{N^2}
 - \frac{6767219}{84000}\frac{\varepsilon ^4}{N^4}  
+ \frac{102957740948201549}{59535676200000}\frac{\varepsilon ^6}{N^6} + O(\varepsilon ^7), \\
&(\rm{k})&  \Delta_1 = \frac{1}{3} + \frac{193}{30}\frac{\varepsilon^2 }{N^2} - \frac{145573959475337}{3308633832000}\frac{\varepsilon ^4}{N^4} + O(\varepsilon ^5), \\
&(\rm{l})&  \Delta_1 = \frac{1}{2} + \frac{23}{6}\frac{\varepsilon^2 }{N^2} + 2\frac{\varepsilon^3 }{N^3}
 - \frac{202777}{7560}\frac{\varepsilon^4 }{N^4} - \frac{72568}{1575}\frac{\varepsilon^5 }{N^5}  + O(\varepsilon ^6),\\
&(\rm{m})&  \Delta_1 = \frac{1}{2} + \frac{593}{106}\frac{\varepsilon^2 }{N^2} + O(\varepsilon ^3),\\
&(\rm{n})&  \Delta_1 = 1 + \frac{4}{3}\frac{\varepsilon^2 }{N^2} - \frac{24289}{4320}\frac{\varepsilon^4 }{N^4} + O(\varepsilon ^5),\\
&(\rm{p})&  \Delta_1 = 2 - \frac{168449}{47505}\frac{\varepsilon^2 }{N^2} + O(\varepsilon ^3),\\
&(\rm{q})&  \Delta_1 = 2 - \frac{47}{15}\frac{\varepsilon^2 }{N^2} + \frac{1}{4}\frac{\varepsilon^3 }{N^3}
 -\frac{241399}{378000}\frac{\varepsilon^4 }{N^4} + \frac{1339}{56700}\frac{\varepsilon^5 }{N^5}+ O(\varepsilon ^6),\\
&(\rm{r})&  \Delta_1 = 3 - \frac{239}{42}\frac{\varepsilon^2 }{N^2} - \frac{168549449269}{41374295424}\frac{\varepsilon ^4}{N^4} + O(\varepsilon ^5),\\
&(\rm{s})&  \Delta_1 = 3 - \frac{239}{42}\frac{\varepsilon^2 }{N^2}
 - \frac{11279063}{3259872}\frac{\varepsilon ^4}{N^4} - \frac{12019873522646189}{6354207699840000}\frac{\varepsilon^6 }{N^6} + O(\varepsilon ^7),\\
&(\rm{t,u})& \Delta_1 = 4 - \frac{719}{90}\frac{\varepsilon^2 }{N^2} - \frac{1658470531}{265356000}\frac{\varepsilon ^4}{N^4}
 \mp \frac{\sqrt{63584702081}}{3302208}\frac{\varepsilon^5 }{N^5} + O(\varepsilon ^6).
\end{eqnarray*}

Asymptotic expansions of boundaries (i') to (u') of the $n:m$ Arnold tongues in Fig.~\ref{tongue_n4}
are
\begin{eqnarray*}
&(\rm{i}')&  \Delta_1 = 2\frac{\varepsilon }{N} + O(\varepsilon ^2) , \\
&(\rm{j'})&  \Delta_1 = \frac{1}{3} + \frac{193}{30}\frac{\varepsilon^2 }{N^2} - \frac{8249819}{84000}\frac{\varepsilon ^4}{N^4} + O(\varepsilon ^5), \\
&(\rm{k'})&  \Delta_1 = \frac{1}{3} + \frac{193}{30}\frac{\varepsilon^2 }{N^2}
 - \frac{6767219}{84000}\frac{\varepsilon ^4}{N^4} + \frac{1286745722182601}{742041300000}\frac{\varepsilon^6 }{N^6} + O(\varepsilon ^7), \\
&(\rm{l'})&  \Delta_1= \frac{1}{2} + \frac{23}{6}\frac{\varepsilon^2 }{N^2} - 2\frac{\varepsilon ^3}{N^3} + O(\varepsilon ^4),\\
&(\rm{m'})&  \Delta_1 = \frac{1}{2} + \frac{23}{6}\frac{\varepsilon^2 }{N^2}+ 2\frac{\varepsilon ^3}{N^3}
 - \frac{202777}{7560}\frac{\varepsilon^4 }{N^4} - \frac{7752}{175}\frac{\varepsilon^5 }{N^5} + O(\varepsilon ^6),\\
&(\rm{n'})&  \Delta_1 = 1 - \frac{2}{3}\frac{\varepsilon^2 }{N^2} + O(\varepsilon ^4),\\
&(\rm{o'})&  \Delta_1 = 1 + \frac{4}{3}\frac{\varepsilon^2 }{N^2} - \frac{17029}{4320}\frac{\varepsilon^4 }{N^4} + O(\varepsilon ^5),\\
&(\rm{p'})&  \Delta_1 = 2 - \frac{47}{15}\frac{\varepsilon^2 }{N^2} - \frac{1}{4}\frac{\varepsilon^3 }{N^3}+ O(\varepsilon ^4),\\
&(\rm{q'})&  \Delta_1 = 2 - \frac{47}{15}\frac{\varepsilon^2 }{N^2} + \frac{1}{4}\frac{\varepsilon^3 }{N^3}
 - \frac{241399}{378000}\frac{\varepsilon^4 }{N^4} + \frac{5027}{6300}\frac{\varepsilon^5 }{N^5}+ O(\varepsilon ^6),\\
&(\rm{r'})&  \Delta_1 = 3 - \frac{239}{42}\frac{\varepsilon^2 }{N^2} - \frac{12094031}{3259872}\frac{\varepsilon ^4}{N^4} + O(\varepsilon ^5),\\
&(\rm{s'})&  \Delta_1 = 3 - \frac{239}{42}\frac{\varepsilon^2 }{N^2} - \frac{11279063}{3259872}\frac{\varepsilon ^4}{N^4}
 - \frac{68218740201013}{50833661598720}\frac{\varepsilon^6 }{N^6} + O(\varepsilon ^7),\\
&(\rm{t',u'})& \Delta_1 = 4 - \frac{719}{90}\frac{\varepsilon^2 }{N^2} - \frac{1658470531}{265356000}\frac{\varepsilon ^4}{N^4}
 \mp \frac{5}{64}\frac{\varepsilon^5 }{N^5} + O(\varepsilon ^6).
\end{eqnarray*}

\end{document}